\pdfoutput=1
\documentclass[apl,twocolumn]{revtex4}
\usepackage{amsmath}
\usepackage{amssymb}
\usepackage{color}
\usepackage{graphics}
\usepackage{graphicx}
\usepackage{dcolumn}

\def\q{\quad}

\begin{document}

\title {Plasmons in graphene on uniaxial substrates}

\author{ I.~Arrazola$^{1,2}$}
\author{R.~Hillenbrand$^{2,3}$}
\author{ A.~Yu.~Nikitin$^{2,3}$}
\email{alexeynik@rambler.ru}
 \affiliation{$^1$ Universidad del Pa\'{i}s Vasco, 48940 Leioa, Spain \\
$^2$ CIC nanoGUNE Consolider, 20018 Donostia-San Sebasti$\mathrm{\acute{a}}$n, Spain\\
$^3$ Ikerbasque, Basque Foundation for Science, 48011 Bilbao, Spain}

\begin{abstract}

Placing graphene on uniaxial substrates may have interesting application potential for graphene-based photonic and optoelectronic devices. Here we analytically derive the dispersion relation for graphene plasmons on uniaxial substrates and discuss their momentum, propagation length and polarization as a function of frequency, propagation direction and  both ordinary and extraordinary dielectric permittivities of the substrate. We find that the plasmons exhibit an anisotropic propagation, yielding radially asymmetric field patterns when a point emitter launches plasmons in the graphene layer.
\end{abstract}


\maketitle

Graphene may open novel avenues for the development of ultra-compact nanoscale photonic and optoelectronic devices\cite{graphphot_natphot10,deAbajoScience}. Particularly interesting applications are expected in optical information processing, where electromagnetic waves have to be guided and external control of their propagation is required. In future graphene-based photonic platforms this may be accomplished by manipulating graphene plasmons (GPs) \cite{Engheta}, which are coherent electron oscillations in doped graphene coupled to an electromagnetic field.  The electromagnetic modes exhibit a wavelength $\lambda_p$ much smaller than that of the free-space, $\lambda_p\ll\lambda$ \cite{Shung86,Vafek06,Wunsch06,Hwang07,Hansonw08,JablanPRB09,Engheta}, which enables them to squeeze into extremely subwavelength volumes. Most importantly, these modes are sensitivie to electrostatic doping of graphene, making them highly attractive for electrical control of light on the nanometer scale. For these reasons, the interest in GPs is steadily increasing in theoretical and experimental communities\cite{GrigorenkoNP12,bludrev}.

Different amazing physical phenomena involving GPs have already been studied, including waveguiding\cite{Engheta,NikitinPRBribbons11,ChristensenACS12,APLsteps12}, strong light-matter interaction\cite{NikitinPRB11,KoppensNL11,StauberPRB11} and enhanced light absorption\cite{NikitinAPL12,ringsAPL12,deAbajoPRL12}.  Recent experimental studies of both propagating and cavity-like graphene plasmons have confirmed their extremely short wavelengths (i.e. high momenta), as well as their remarkable tunability by electrical gating \cite{PlasmonicsNature11,NatureDiscs12,KoppensNature12,BasovNature12,IBMNaturePhot13}.

A deposition of graphene onto different metamaterial (anisotropic) substrates\cite{ZheludevOE10} could pave the way to promising hybrid optical elements and nanodevices includng biochemical sensors, polarizers and modulators \cite{GrigorenkoNP12}. However, the impact of an anisotropic material on intrinsic plasmons in graphene has not been addressed yet. In this letter we introduce a concept for manipulating the propagation of GPs with the use of anisotropic substrates. As a proof of concept, we consider the case of a uniaxial substrate. Uniaxial crystals can be regarded as the long-wavelength limit of media composed by periodically-stacked layers (one-dimensional photonic crystals)\cite{Yariv}, as illustrated in Fig. 1. Uniaxial crystals as for example SiC or heteroepitaxial structures represent an important family of anisotropic media with a wide range of applications from photonics to electronics\cite{OcelicOE08,Heteroepitaxial}.

In order to treat a periodic layered substrate as a uniaxial crystal, the lattice constant $a$ has to be much smaller than the GP wavelength, $a\ll\lambda_p$. For example, in the mid infrared (IR) frequency region, where $\lambda_p$ is of order of the wavelength of the visible light, the lattice period should fulfill the same requirements as photonic crystals in the visible spectral range, i.e. the lattice constant should raughty satisfy $a<100$ nm \cite{TornerPRL05}.

\begin{figure}[thb!]
\includegraphics[width=7cm]{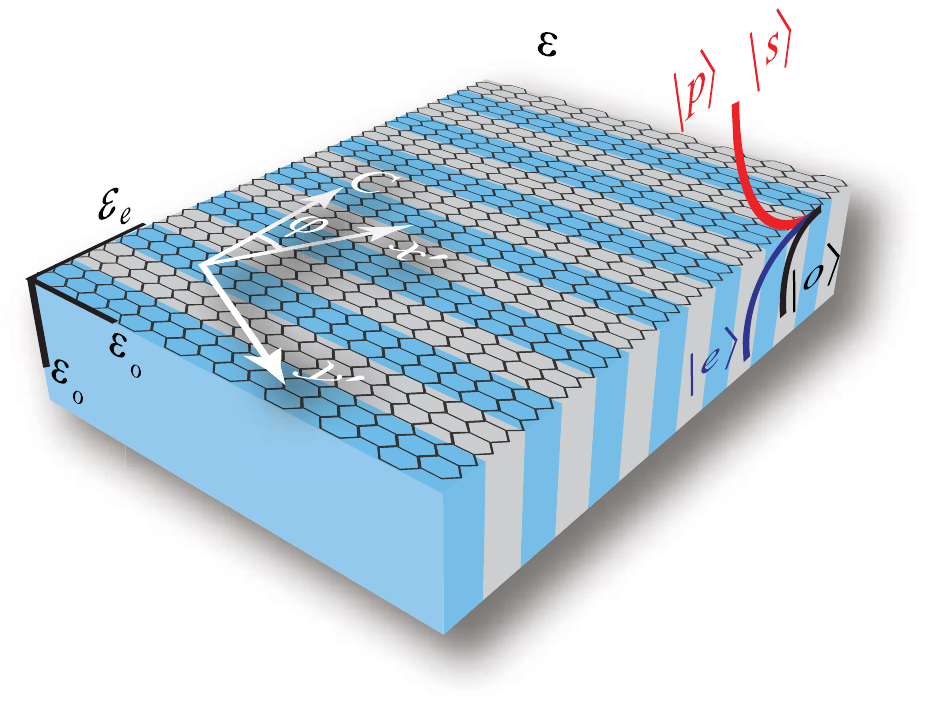}\\
\caption{(Color online)  Schematic of the studied system. A graphene monolayer is placed onto the surface of a uniaxial crystal with ordinary and extraordinary dielectric permittivities $\varepsilon_o$ and $\varepsilon_e$. The upper homogeneous half-space has the dielectric permittivity $\varepsilon$. The plasmon field is composed by a superposition of the eigenmodes of the substrate and upper half-space. $|s>$ and $|p>$ present  the $s$- and $p$-polarized waves in the upper half space (with the same decay), while $|o>$ and $|e>$ correspond to the ordinary and extraordinary waves in the uniaxial substrate (decaying differently).}\label{geom}
\end{figure}

A schematic of the system studied in this letter is presented in Fig. 1. A graphene monolayer is placed onto a uniaxial substrate with the dielectric tensor $\hat{\varepsilon}$ and covered by an upper homogeneous half-space with the homogeneous dielectric permittivity $\varepsilon$. We assume the surface of the uniaxial substrate to be parallel to two principal axes so that the dielectric tensor $\hat{\varepsilon}$ has the following form ($x$-axis coincides with the main axis $C$)
\begin{equation}\label{eps}
\hat{\varepsilon} =
\begin{pmatrix}
\varepsilon_{e} & 0 & 0\\
0 & \varepsilon_{o} & 0\\
0 & 0 & \varepsilon_{o}\\
\end{pmatrix}.
\end{equation}
The graphene layer is described by its two-dimensional conductivity $\sigma$, which is obtained within the random phase approximation\cite{Wunsch06,Hwang07,Falkovsky08}. In the following we look for a solution of Maxwell's equations in the form of an electromagnetic wave which field is exponentially decaying perpendicular to the graphene sheet. In order to simplify the representation of the results and calculations, we use a rotated coordinate system $(x',y',z)$ in which the wave propagates along the $x'$-axis forming the angle $\varphi$ with the main crystal axis $C$ (see Fig.~1). The fields in both the uniaxial substrate and the upper half-space are projected onto linearly independent modes. In the uniaxial substrate they are denoted as $|o>$ and $|e>$, representing ordinary and extraordinary electromagnetic waves respectively. In the upper half-space they are denoted $|s>$ and $|p>$, corresponding to $s$- and $p$-polarized plane waves, respectively. Each mode is given by the product of its polarization vector $\mathbf{e}_i$ and the oscillating function $\exp(ig\mathbf{Q}_j\mathbf{r})$, where  $\mathbf{r}=(x',y',z)$, $j=s,p,o,e$ is the position vector and $g=2\pi/\lambda$ the vacuum wavevector. The wavevectors for all plane waves are given by $\mathbf{Q}_j=(q,0,\pm q_{zj})$, where the signs $+$ and $-$ correspond to birefringent and homogeneous half-space, respectively.  The polarization vectors are given by
\begin{equation}\label{e}
\begin{split}
&\mathbf{e}_s = \mathbf{e}_y, \q \mathbf{e}_p = \mathbf{e}_s\times\frac{\mathbf{Q}_s}{Q_s},\\
&\mathbf{e}_e = \varepsilon_o\mathbf{e}_x-(\mathbf{e}_x\cdot\mathbf{Q}_e)\mathbf{Q}_e, \q \mathbf{e}_o = \mathbf{e}_C\times \mathbf{Q}_o,
\end{split}
\end{equation}
and the $z$-components of the wavevectors satisfy the following dispersion relations
\begin{equation}\label{kz}
\begin{split}
q^2_{z} + q^2 =\varepsilon, \q q^2 + q^2_{zo}=\varepsilon_o,\q q^2_{ze} + q^2\beta^2(\varphi) =\varepsilon_e,
\end{split}
\end{equation}
where $q_{zs}=q_{zp}=q_{z}$ and $\beta^2(\varphi) = \sin^2\varphi+(\varepsilon_e/\varepsilon_o)\cos^2\varphi$. Because of the radiation conditions it is required that $\mathrm{Im}(q_{zj})\geq0$ for all $j$. By matching the fields between the upper half-space ($\mathbf{E}_1$ and $\mathbf{H}_1$) and the substrate ($\mathbf{E}_2$ and $\mathbf{H}_2$) according to the boundary conditions (i.e. continuity of the parallel components of the electric fields, $\mathbf{e}_z\times(\mathbf{E}_1-\mathbf{E}_2)=0$ and discontinuity of the parallel components of the magnetic fields due to the presence of the conducting layer, $\mathbf{e}_z\times(\mathbf{H}_1-\mathbf{H}_2) =  2\alpha\,\mathbf{e}_z\times(\mathbf{e}_z\times\mathbf{E}_1)$),
we obtain the following dispersion relation
\begin{equation}\label{disp}
\begin{split}
&q_{zo}\left(q_z +2\alpha + q_{zo}\right)\left[q_zq_{ze}\varepsilon_o +q^2_{zo}(\varepsilon+2\alpha q_z)\right]\cos^2\varphi \\
&+\varepsilon_o\left(q_z+2\alpha+q_{ze}\right)\left[q_{zo}(\varepsilon+2\alpha q_z)+  q_z\varepsilon_o \right]\sin^2\varphi=0.
\end{split}
\end{equation}
where $\alpha$ is the normalized conductivity, $\alpha=2\pi \sigma /c$. Eq.~\eqref{disp} describes different types of surface waves and has been studied in some limiting cases. For instance, if the graphene is not present (i.e. $\alpha=0$), Eq.~\eqref{disp} corresponds to Dyakonov waves under the condition $\varepsilon_e>\varepsilon>\varepsilon_o$ \cite{Dyakonov88,TornerPRL05,NikitinOL09}, or hybrid surface plasmons if the upper half-space is metal, i.e. $\mathrm{Re}(\varepsilon)<0$ \cite{anizsppAPL08}. In the case that graphene is placed onto a homogeneous substrate with the same dielectric permittivity as the upper half-space ($\varepsilon_o = \varepsilon_e = \varepsilon$), Eq.~\eqref{disp} describes plasmons in a symmetrically-surrounded graphene sheet. The dispersion relation in this case has two linearly independent solutions: GPs, $q_z\alpha+\varepsilon=0$, and $s$-polarized surface waves on graphene, $q_z+\alpha=0$ \cite{Hansonw08}.

\begin{figure}[thb!]
\includegraphics[width=6cm]{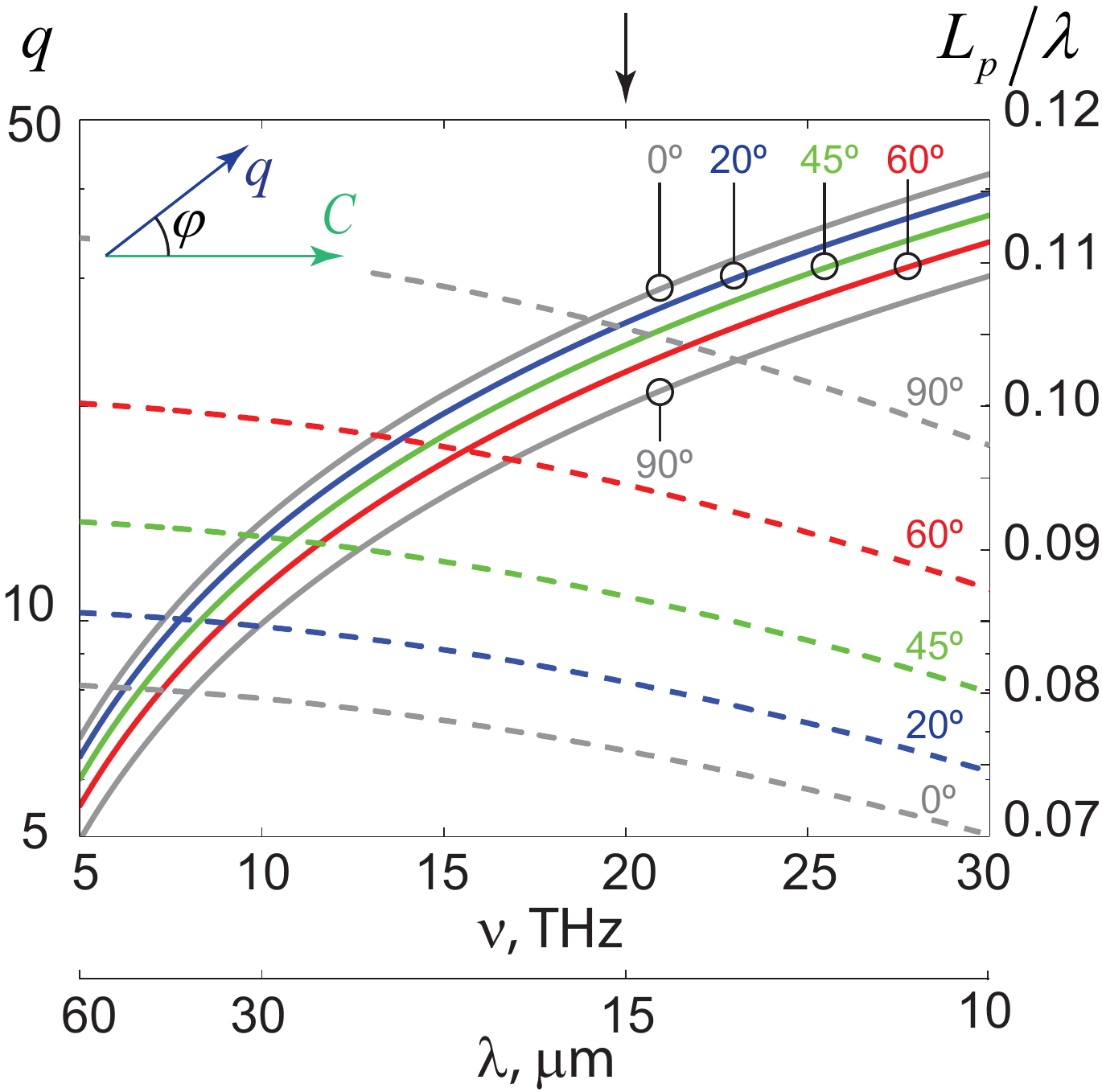}\\
\caption{(Color online) GPs momentum $q$ (continuous curves) and propagation length in units of the wavelength $L_p/\lambda$ (discontinuous curves) as a function of frequency/wavelength at different values of $\varphi$. Crystal permittivities are $\varepsilon_o=2$ and $\varepsilon_e=5$, while the bounding medium is vacuum, $\varepsilon=1$. The parameters of the graphene are: Fermi level $E_F=0.4$ eV, relaxation time $\tau = 0.1$ ps, temperature $T=300$ K.}\label{qom}
\end{figure}

In order to analyse GPs on an arbitrary uniaxial substrates, Eq.~\eqref{disp} generally needs be solved numerically. However, it can be greatly simplified in the mid-IR spectral range, where the GP momentum is up to two orders of magnitude larger than the momentum of light in free space\cite{KoppensNature12,BasovNature12}.  Assuming that $q\gg \mathrm{max(}\sqrt{\varepsilon},\sqrt{\varepsilon_o},\sqrt{\varepsilon_e})$, we obtain the following expression from Eq.~\eqref{disp}:
\begin{equation}\label{dispa}
\begin{split}
q = q(\omega,\varphi) = \frac{\pm\beta(\varphi)\varepsilon_o+\varepsilon}{2}q_0(\omega),
\end{split}
\end{equation}
where $q_0(\omega)\simeq i/\alpha$ is the GP wavevector for a free-standing graphene (the argument $\omega$ will be omitted hereafter). The choice of the sign in Eq.~\eqref{dispa} is dictated by the need of positive imaginary parts of both $x$- and $z$-components of all the wavevectors. We clearly see that $q(\varphi)$ is the product of the GP wavevector for a free-standing graphene layer $q_0$ and the function $\frac{\pm\beta(\varphi)\varepsilon_o+\varepsilon}{2}$, which depends on the propagation direction and dielectric permittivities of the uniaxial substrate. It directly follows that the GP propagation length $L_{p}=\lambda/[2\pi \mathrm{Im}(q)]$ expressed in units of the GP wavelength $\lambda_p=\lambda/[2\pi \mathrm{Re}(q)]$ is independent on $\varphi$, and is only dependent on the intrinsic properties of graphene, $L_{p}/\lambda_p \simeq \mathrm{Im}(\alpha)/\mathrm{Re}(\alpha)$ so that the GP decays within the same number of the oscillation periods in any direction.

\begin{figure}[thb!]
\includegraphics[width=6cm]{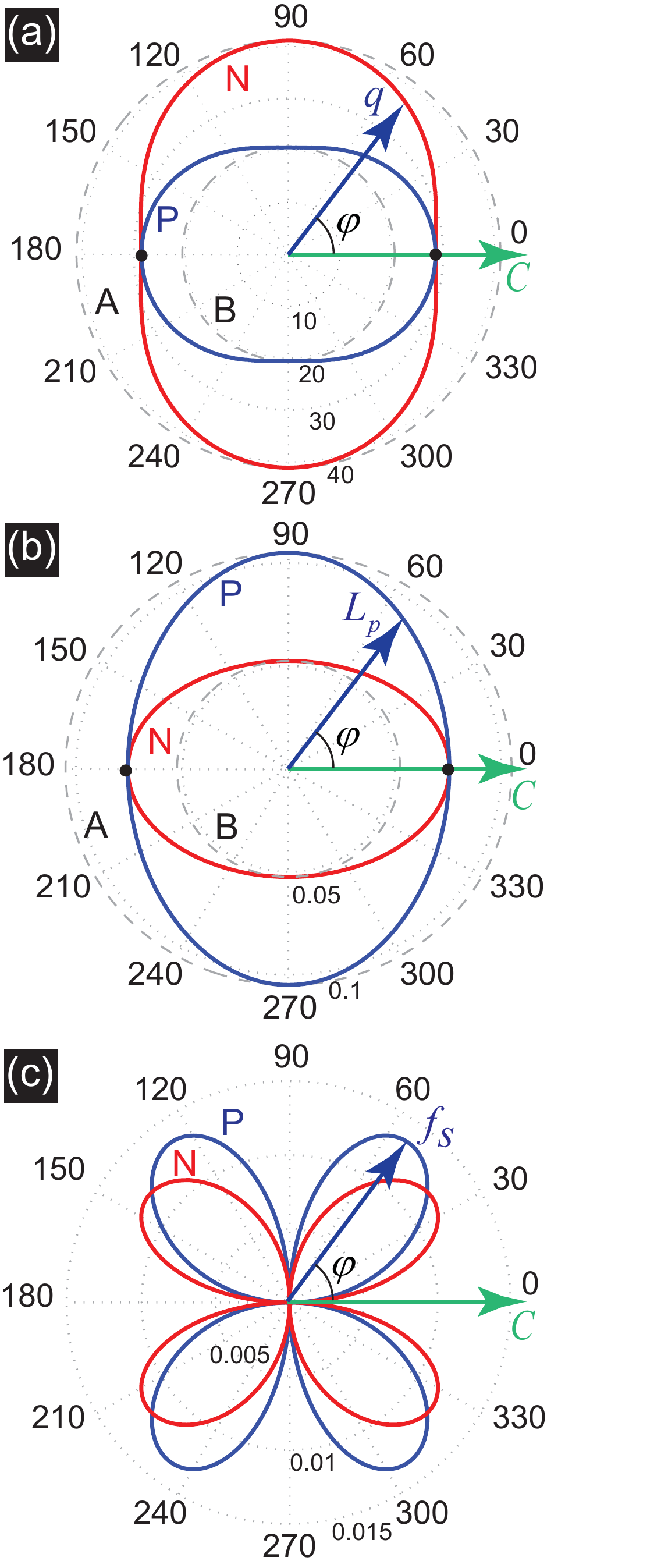}\\
\caption{(Color online) GPs momentum $q$ in (a), propagation length in units of the wavelength $L_p/\lambda$ in (b) and polarization ratio $f_s$ in (c) as a function of the angle $\varphi$ for the frequency of 20 THz (corresponding to the black arrow on top of Fig.~2). In all 3 panels the curves ``P'' and ``N'' correspond to the positive crystal with $\varepsilon_o=2$, $\varepsilon_e=5$ and negative one with $\varepsilon_o=5$, $\varepsilon_e=2$, respectively; $\varepsilon=1$. The black discontinuous circles in ``A'', ``B'' in (a) and (b) represent the asymptotic values of $q$ and $L_p/\lambda$ for GPs in a homogeneous substrate with $\varepsilon_h$. In A $\varepsilon_h=5$ while in B $\varepsilon_h=2$. The parameters for the graphene are the same as in Fig.~2.}\label{polar}
\end{figure}

In the following we discuss the dispersion of GPs on a substrate where the dielectric values $\varepsilon_o$ and $\varepsilon_e$ for ordinary and extraordinary directions are real and positive. The upper half space is vacuum, $\varepsilon=1$. We consider both positive ($\varepsilon_e>\varepsilon_o$) and negative ($\varepsilon_e<\varepsilon_o$) uniaxial crystals. For illustrative purposes we take $\varepsilon_o=2$, $\varepsilon_e=5$ for the positive crystal and $\varepsilon_o=5$, $\varepsilon_e=2$ for the negative one. For a multilayer made of two media with dielectric permittivities $\varepsilon_1$ and $\varepsilon_2$, and filling factor $\eta$, the considered case of the positive crystal corresponds for instance to $\varepsilon_1=-100$, $\varepsilon_2=4.9$, $\eta=35.7$, while in case of the negative one $\varepsilon_1=1$, $\varepsilon_2=8$, $\eta=1.3$ (according to effective medium approximation \cite{Yariv}). Note that the dielectric metamaterial crystal could be created as a thin slab and placed on a silicon substrate. The silicon could be used as a backgate to electrostatically dope the graphene, similarly to experiments where graphene is placed on a SiO$_2$ layer on a silicon substrate\cite{KoppensNature12,BasovNature12}.

Fig.~2 shows the wave vector of the GPs, $q$, and the propagation length, $L_p$, as a function of the frequency $\nu$ for different propagation directions $\varphi$. For calculating the graphene conductivity we assume a Fermi level $E_F=0.4$. Such Fermi levels can be achieved by intrinsic graphene doping or by electrostatic gating of the graphene described above\cite{KoppensNature12,BasovNature12}. The temperature is set to $T=300$K and for the relaxation time we assume $\tau = 0.1$ ps (corresponding to a mobility of 2500 $\mathrm{cm^2/(Vs)}$). We use the Eq.~\ref{dispa}, although in the shown frequency region the result is undistinguishable from the numerical solution of the Eq.~\ref{disp}. Similarly to GPs on isotropic substrates, the GP wavevector increases with frequency, while its propagation length decreases \cite{Hansonw08,JablanPRB09}. The set of the dispersion curves for different values of $\varphi$ is restricted between the asymptotic values $q(0) = \frac{\sqrt{\varepsilon_o\varepsilon_e}+\varepsilon}{2}q_0$ and $q(\pi/2) = \frac{\varepsilon_o+\varepsilon}{2}q_0$. Remarkably, the latter asymptote $q(\pi/2)$ presents the GPs wavevector for graphene on a homogeneous substrate with the permittivity $\varepsilon_o$. Notice that the cases of $\varphi$, $-\varphi$, $\pi-\varphi$, and $\pi+\varphi$ are equivalent, and thus in Fig.~2 we only show the range of $0<\varphi<\pi/2$.

\begin{figure}[thb!]
\includegraphics[width=7cm]{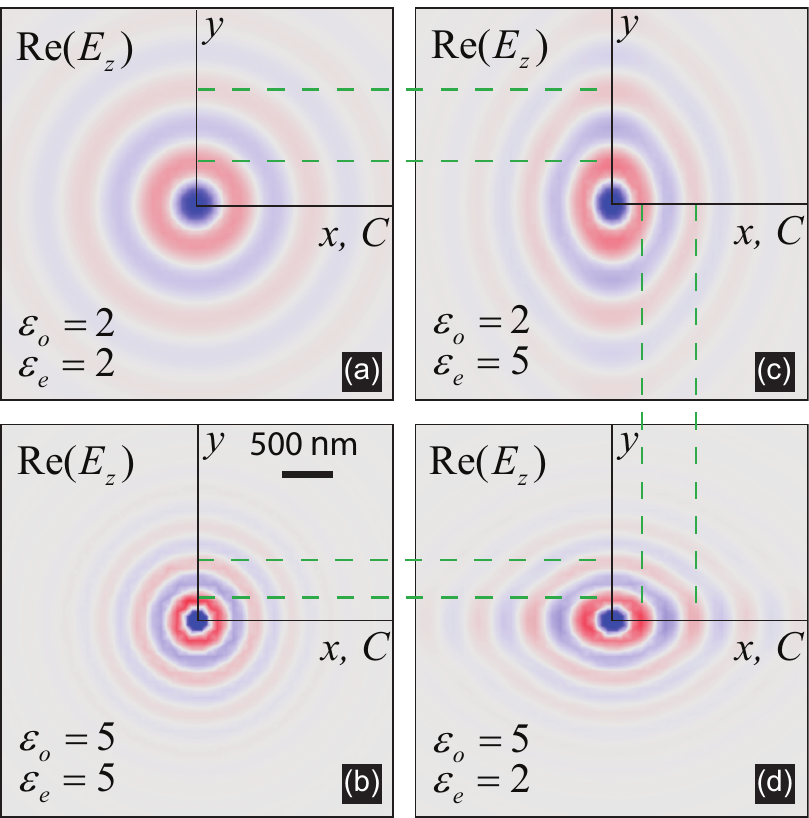}\\
\caption{(Color online) Instant snapshots of the $z$-component of the electric field along graphene sheet created by a vertical point dipole. The dipole is placed at the height of 25 nm from the monolayer on different substrates.  In (a) and (b) the substrates are homogeneous, while (c) and (d) they present positive and negative crystals respectively. The parameters for the graphene are the same as in Fig.~2, the frequency is 20 THz. The vertical and horizontal discontinuous lines are guides to the eyes.}\label{dipole}
\end{figure}

In order to better illustrate the anisotropy of the GP dispersion we render in Fig.~3(a) the wavevector $q$ as a function of $\varphi$ at a fixed frequency. In case of the negative crystal, $q(\varphi)$ has an elongated form in the direction perpendicular to the main crystal axis $C$, while for the positive crystal $q(\varphi)$ is elongated along $C$. At $\varphi=\pi/2$ the dispersion curves reach their asymptotic values $q(\pi/2)$ (shown by the discontinuous circles in Fig.~3(a)). Interestingly, for the negative crystal $q(\varphi)<q(\pi/2)$, while for the positive one $q(\varphi)>q(\pi/2)$. Another point to note is that due to the symmetry of $q(0)$ with respect to the change between $\varepsilon_o$ and $\varepsilon_e$, the dispersion curves for negative and positive crystals coincide at $\varphi=0$ (this is shown by the black points in Fig.~2(a),(b)). This means that for  GPs propagating  along the axis $C$, negative and positive crystals are identical.

An important characteristic of GPs is their propagation distance $L_p$ which is represented in Fig.~3(b) as a function of $\varphi$ in units of $\lambda$. It is easy to see that the curves $L_p(\varphi)$ and $q(\varphi)$ corresponding to the same crystals  (compare panels (a) and (b) in Fig.3) are elongated in the opposite directions. This is related to the fact that for higher values of momentums the confinement of GPs is higher and they are absorbed at shorter absolute distances.

Additionally, we complement the dispersion of GPs by studying the polarization. GPs on an anisotropic substrate are hybrid waves which is necessary to meet the matching conditions on the interface between the fields in the crystal and homogeneous region. The dependency of ratio between the $s$- and $p$-polarized components $f_s = |E_s/E_p|$ on $\varphi$ is shown in Fig.3(c). In the whole angular interval $f_s(\varphi)$ has four lobes which are due to the following proportionality $f_s(\varphi)\propto \sin2\varphi$. In the shown case the portion of the $s$-component is below 1.5 $\%$, while in the whole frequency region shown in Fig.~2 it does not exceed 12 $\%$.

Finally, to corroborate the analytical solution for the GPs dispersion relation, we study the excitation of GPs by a point source (see Fig.~4). We conduct finite-elements calculations\cite{COMSOL} of the fields created by a point dipole placed at a short distance (25 nm) from the graphene sheet. The dipole moment is parallel to $z$-axis so that the field pattern for a homogeneous substrate $\varepsilon_o=\varepsilon_e=\varepsilon_h$ are radially symmetric\cite{NikitinPRB11,KoppensNL11} (see Fig.~4(a,b)). The period of the near-field oscillations (given by the GP wavelength $\lambda_p$) observed in the snapshots for a homogeneous substrate at a fixed $\lambda$ is constant in any direction from the dipole. It depends on the permittivity of the substrate $\lambda_p=2\lambda_{p0}(\varepsilon_h+1)^{-1}$ with $\lambda_{p0}$ being the GP wavelength for a free-standing graphene, so that for larger $\varepsilon_h$ the GP wavelength is shorter (compare panels (a) and (b) in Fig.~4). In contrast, for the uniaxial substrate the separation between the field maxima is dependent on the propagation direction, $\lambda_p=\lambda_p(\varphi)=\lambda/[2\pi \mathrm{Re}(q(\varphi))]$ (see Fig.~4(c,d)). The snapshots presented in Fig.~4 confirm our results obtained from the analytical solutions of the dispersion relation (Figs. 2 and 3). Indeed, along the axis $C$, $\lambda_p$ is the same for the positive and negative crystals, being in the shown case $\lambda_p=503$ nm which is in correspondence with $q= 29.8$ given by Eq.~\eqref{dispa}. In the perpendicular direction (along the $y$-axis), both for the positive and negative crystals $\lambda_p$ coincides with the one for a homogeneous substrate with $\varepsilon_h=\varepsilon_o$ so that in Fig.~4(a) $\lambda_p=679$ nm while in Fig.~4(b) $\lambda_p=340$ nm. This is also in accordance with Eq.~\eqref{dispa} which yields for these two cases $q=21.5$ and $q=43$ respectively. For a better visualization of the above findings, the vertical and horizontal discontinuous lines are drawn in Fig.~4.

In summary, in this letter we provide an analytical solution for the dispersion relation of GPs on a uniaxial substrate. We have shown that the momentum of the GPs is dependent on the propagation direction. When deviating from the main crystal axis $C$, the momentum of the GP increases for negative and decreases for positive crystals, respectively. If, however, GPs propagate along the C-axis, the GP momentum is the same for negative and positive crystals and equals that of the GP momentum on a homogeneous substrate with $\varepsilon_h=\varepsilon_o$. In certain directions GPs acquire the $s$-polarization component, becoming thus hybrid waves. As a consequence of the anisotropy of the GPs propagation, the plasmon field patterns created by a point source are not circularly symmetric, i.e. the distance between the field maxima depends on the propagation direction. This can find interesting applications in the design of graphene plasmonic circuits, in which the resonant wavelengths of the cavities and waveguides depends on their orientation.


\begin{thebibliography}{99}
\bibitem{COMSOL} A commercial software \textsc{comsol} has been used. Graphene has been simulated as a two-dimensional conductivity in the boundary conditions.
\end{thebibliography}
\end{document}